\documentclass[a4paper, prl, twocolumn, showpacs]{revtex4}

\usepackage{amssymb}
\usepackage{amsmath}
\usepackage{epsfig}
\usepackage{color}
\usepackage{graphics, graphicx}
\usepackage{bbold}

\begin{document}

\author{Michiel Snoek}
\author{Masudul Haque}
\author{S. Vandoren}
\author{H. T. C. Stoof}
\affiliation{Institute for Theoretical Physics, Utrecht University, Leuvenlaan 4, 3584 CE Utrecht, The
Netherlands}
\date{\today}

\title{Ultracold Superstrings in atomic Boson-Fermion mixtures}

\begin{abstract}
We propose a setup with ultracold atomic gases that can be used to make a nonrelativistic superstring in four spacetime dimensions. In particular, we consider for the creation of the superstring a fermionic atomic gas that is trapped in the core of a vortex in a Bose-Einstein condensate. We explain the required tuning of experimental parameters to achieve supersymmetry between the fermionic atoms and the bosonic modes describing the oscillations in the vortex position. Furthermore, we discuss the experimental consequences of supersymmetry.
\end{abstract}

\maketitle

In recent years, three topics have in particular attracted a lot of attention in the area of ultracold atomic gases.
These topics are vortices \cite{ref3,ref4,ref5,ref6, ref12}, Boson-Fermion mixtures
\cite{ref17, ref18, ref19, ref20, ref22}, and optical lattices
\cite{ref9, ref10}. 
In this Letter we propose to combine these three topics to engineer a superstring in the laboratory, i.e., a line-like quantum object with both bosonic and fermionic excitations and a supersymmetric hamiltonian that is invariant under interchanges of these excitations.
The physics of a vortex line in a one-dimensional optical lattice has been studied recently \cite{ref7, ref25}. 
Because of the optical lattice, the transverse quantum fluctuations of the vortex line are greatly enhanced in this configuration. The vortex can therefore be viewed as a quantum mechanical string and it forms the bosonic part of our superstring.
In addition, we propose to trap fermionic atoms in the vortex core, to make a  nonrelativistic version of a so-called Green-Schwarz superstring in four spacetime dimensions.  Because our ultracold superstring is nonrelativistic, it is not constrained to the ten-dimensional spacetimes in which superstrings are usually studied in high-energy physics.  The precise mathematical connection with string theory is currently under investigation.

Apart from the connection to string theory, our ultracold superstring  is also of interest in its own right. To the best of our knowledge, it is the 
first condensed-matter system proposed, where supersymmetry can be studied experimentally. The intruiging possibility of observing effects of supersymmetry is common in high-energy physics, but novel in a condensed-matter setting. In this particular case, the supersymmetry protects the superstring from spiraling out of the gas.
This can be understood from the fact that the dissipation resulting in this motion has two sources, namely the creation of two additional bosons in the transverse oscillations of the vortex, and the production of an additional particle-hole pair of fermions. At the supersymmetric point these two contributions interfere destructively and the stability of the superstring is greatly enhanced.
Moreover, the ultracold superstring allows for the study of a quantum phase transition that spontaneously breaks supersymmetry.
Experimentally this will be directly visible by observing the superstring spiraling out of the center of the gas. Note that supersymmetry can only be realized at zero temperature and we consider only this case from now on.

To form the superstring we start with a cigar-shaped Bose-Einstein condensate. The symmetry axis will be called the $z$-axis from now on. Rotation of the condensate along the $z$-axis creates a vortex. Above a critical external rotation frequency $\Omega_c$ a vortex in the center of the condensate is stable. For $\Omega < \Omega_c$ the vortex is unstable, but because of its Euler dynamics, it takes a relatively long time before it spirals out of the gas \cite{ref12, ref13, ref14}. We analyze in detail the case of $\Omega=0$, i.e., the situation in which the condensate is no longer rotated externally after a vortex is created. However, the physics is very similar for all $\Omega < \Omega_c$, where supersymmetry is possible.

Next a one-dimensional optical lattice is imposed. Such an optical lattice consists of two identical counter-propagating laser beams and provides a periodic potential for atoms. When applied along the $z$-axis of the condensate,  the optical lattice divides the condensate into weakly-coupled pancake-shaped condensates, each containing typically of the order of $N_B \simeq 10^3$ bosonic atoms. Moreover, in the case of a red-detuned lattice, the gaussian profile of the laser beam provides also the desired trapping in the radial direction.  
In the one-dimensional optical lattice the vortex line becomes a chain of so-called pancake vortices.
Since the quantum fluctuations of the vortex position are proportional to $1/N_B$ \cite{ref7}, they are greatly enhanced in this configuration as compared to the bulk  situation. An added advantage of the stacked-pancake configuration is the particle-like dispersion of the vortex oscillations, which ultimately allows for supersymmetry with the fermionic atoms in the mixture.
Nearest-neighbor pancake vortices attract each other due to the Josephson effect that is a result of the hopping of atoms between neighboring wells of the optical lattice. The stiffness of the vortex line is therefore determined by the hopping amplitude. The vortex oscillations are bosonic excitations with a tight-binding dispersion \cite{ref7}. These excitations are known as Kelvin modes and provide the bosonic part of the superstring. Without the optical lattice Kelvin modes have already been observed experimentally \cite{ref15,ref16}. 

For the fermionic degrees of freedom of the superstring, we need to place a fermionic atomic species in the vortex core. If the interspecies interaction between fermionic and bosonic atoms is repulsive, the fermions can be made to populate bound states in the vortex core where the boson density is lowest. More precisely, the ratio $a_{BF}/a_{BB}$ of the boson-fermion scattering length and the boson-boson scattering length needs to be sufficiently large. 
The trapping of atoms in bound states within a vortex core has already been achieved for atoms in a different hyperfine state \cite{ref4, ref12}. 
The presence of fermions within the vortex core in principle provides a mass to the vortex. This modifies the Euler dynamics, but this effect can safely be ignored under all conditions of interest to us, because the ratio $N_F/N_B$ of the number of fermionic atoms and the number of bosonic atoms per site turns out to be very small.  The general properties of our proposed system are summarized in Fig. 1.

\begin{figure}
\includegraphics[scale=1.0]{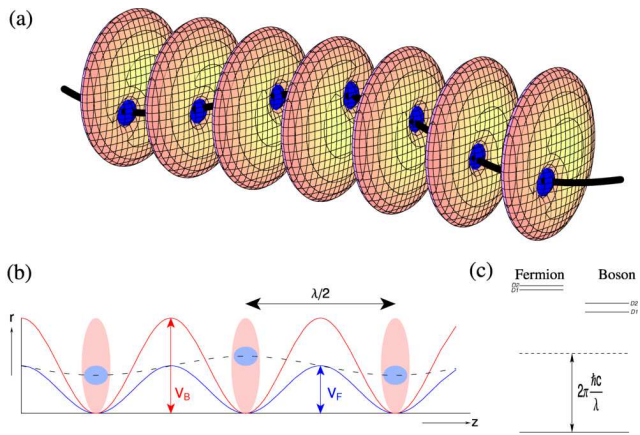}
\caption{ (color online) (a) Artist's impression of the setup. The disks represent the bosonic condensate density and the blue balls in the vortex core represent the fermionic density. The black line is a guide to the eye to see the wiggling of the vortex line that corresponds to a Kelvin mode. (b) Schematic of the setup. Here $r$ is the radial distance in the $xy$-plane. The pink and blue blobs represent the bosonic and fermionic densities, respectively. Moreover, $\lambda$ is the wavelength of the laser. The blue and red lines indicate the strength of the optical potential, respectively, for the bosons and fermions as a function of the $z$-coordinate. (c) Schematic fine structure level scheme of the bosonic and fermionic atomic species. Because we consider only sufficiently large detunings the hyperfine level structure is not resolved.}
\end{figure}

A convenient choice for the boson-fermion pair is $^{87}$Rb and $^{40}$K, since such Boson-Fermion mixtures have recently been realized in the laboratory \cite{ref17, ref18, ref19, ref20} and because the resonance lines in these two atomic species lie very nearby. In this case, for typical bosonic densities, the ratio $a_{BF}/a_{BB}$ needs to be larger than 2 to have a bound state in the vortex core. This is calculated by using the condensate density profile as an effective potential for the fermions. The interspecies scattering length can be made positive and large either by choosing the appropriate spin states or by means of various broad Feshbach resonances that can make the interaction repulsive while keeping the probability to create molecules negligible \cite{ref21}. The calculations presented in this Letter are for the $^{87}$Rb - $^{40}$K mixture, but in principle it is also possible to use other mixtures. Another Boson-Fermion mixture that has been realized in the laboratory consists of $^{23}$Na and $^{6}$Li atoms \cite{ref22}. This mixture is less convenient because the resonance lines are widely separated, so that the two species feel very different optical potentials and it is hard to trap both with a single laser. In addition, $^6$Li is relatively hard to trap in an optical lattice because of its small mass. For these reasons, the $^{23}$Na-$^6$Li mixture can only be used in a very restricted parameter regime, as shown later on in Fig. 2.

The laser intensity needs to be strong enough that there is a bound state for the atoms at each site, but not strong enough to drive the system into a Mott insulator state \cite{ref9, ref10, ref11}, in which case the pancake vortices no longer bind together.
Given the frequencies of the $D_1$ and $D_2$ resonance lines $\omega_{D_1}$ and $\omega_{D_2}$, respectively, the optical potential for the atoms is given by $V_{B,F} (z) =  V_{B,F} \cos^2[2\pi z/\lambda]$, where the well depths obey
\begin{eqnarray*}
V_{B,F}&=&\hbar \Omega_{B,F}^2 \left[
\frac{1}{3} \left(  \frac{1}{\omega^{B,F}_{D_1} - \omega}
+ \frac{1}{\omega^{B,F}_{D_1} + \omega} \right) \right. \\
&&\hspace{1cm} \left.+\frac{2}{3}
\left( \frac{1}{\omega^{B,F}_{D_2} - \omega} +
\frac{1}{\omega^{B,F}_{D_2} + \omega}
\right) \right], \nonumber
\end{eqnarray*}
$\omega= 2 \pi c/\lambda$ is the laser frequency, and $\Omega_{B,F}$ are the Rabi frequencies for the bosonic and fermionic atoms in the mixture, respectively.

Because of the use of a single laser the fermionic and bosonic atoms experience lattice potentials with the same periodicity, but with different well depths (cf. Fig. 1b). Ignoring their interactions, the fermionic atoms and the kelvon excitations now both have a tight-binding dispersion $2 J [1- \cos (k \lambda /2) ]$, with hopping amplitudes $J_F$ and $J_K$, respectively. Therefore, by making the hopping amplitudes of the kelvons and the fermionic atoms equal, the lattice laser can be tuned in such a way that the kelvon and fermion dispersions coincide, which is a first
requirement for supersymmetry. 
For a sufficiently deep potential the atomic hopping amplitudes are given by the exact result
$
J_{B,F} = 4 (V_{B,F})^{3/4} (E_{B,F})^{1/4} \exp\left[-2
\sqrt{V_{B,F}/{E_{B,F}}}\right]/\sqrt{\pi},
$
where $E_{B,F}= 2 \pi^2 \hbar^2 /m_{B,F} \lambda^2$ are the recoil energies and $m_{B,F}$ are the atomic masses. The kelvon hopping amplitude is given by $J_K=\Gamma[0, (\ell/R)^4]J_B$, where $\ell$ is the radial harmonic-oscillator length for the bosonic atoms, $R$ is the Thomas-Fermi radius of every pancake, and $\Gamma[0, z]$ is the incomplete gamma function \cite{ref7}. The ratio $\ell/R$ is determined by the number of bosonic atoms per site $N_B$ and the strength of their repulsive interaction. One criterion for supersymmetry is, therefore,
\begin{equation}
\Gamma[0, (\ell/R)^4] J_B = J_F \equiv t.
\end{equation}
In Fig. 2 is indicated how to tune the wavelength and Rabi frequency of the lattice laser to achieve this. Note that we implicitly assumed that the fermionic hopping parameter is independent of the presence of kelvons, which is justified for the long wavelengths of interest, because then the difference between the positions of the nearest-neighbor pancake vortices is negligible with respect to the core size.

\begin{figure}
\includegraphics[scale=.63]{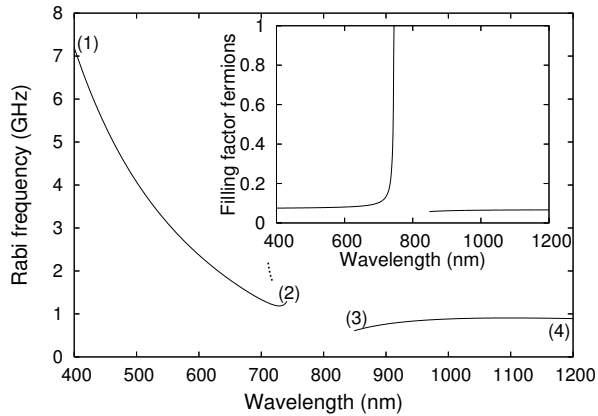}
\caption{Tuning of the lattice laser to obtain supersymmetry for 1000 bosonic atoms per lattice site: Rabi frequency for the bosonic atoms versus wavelength for $^{87}$Rb-$^{40}$K (full line) and $^{23}$Na-$^6$Li (dotted line). Note that for the blue-detuned part, i.e., $\lambda < 760$ nm for the  $^{87}$Rb-$^{40}$K mixture extra radial trapping is needed, either magnetically, or by using an extra running laser as discussed later in the text and shown in Fig. 3. In Fig. 3 we display how to tune the running laser to obtain also supersymmetric interactions. The numbers (1)-(4) indicate the parameters for which this is calculated. The inset shows the average number of fermions per lattice site. This depends linearly on the ratio of the harmonic lengths in the axial and radial directions. This ratio should be sufficiently small to be radially in the Thomas-Fermi limit. For this plot a ratio of 1/5 is chosen.}
\end{figure}

A second requirement is that the chemical potential of the kelvons is equal to the chemical potential of the  fermions. Therefore, the filling factor of the fermions has to be adjusted accordingly.
Equating the chemical potential of the kelvons and fermions we get
\begin{equation}
\mu_K = \mu_F = \frac{\hbar^2 }{2  m_B R^2}(1-\Gamma[0,
(\ell/R)^4]) \equiv \mu.
\end{equation}
From this we can derive that the average number of fermions per lattice site is given by $N_F = 2 \arcsin[\sqrt{\mu_F/4 J_F}]/\pi$.
For typical parameters the number of fermions per lattice site turns out to be of the order of 0.1, as shown in  the inset of Fig. 2. Tuning the system parameters to lie on the curves in Fig. 2 makes the string supersymmetric.  The system is then invariant under unitary transformations of the bosonic and fermionic excitations among each other.

In our superstring realization there are also boson-boson and boson-fermion interactions. The kelvons interact repulsively among each other when $\Omega <  \Omega_c$ \cite{ref25}. In addition, a repulsive interaction between the kelvons and the fermionic atoms is generated by the fact that physically the presence of a kelvon means that the vortex core moves off center, together with the fermions trapped in it. Because of the confinement experienced by the trapped fermions in the radial direction, this increases the energy of the vortex. The kelvon-kelvon interaction coefficient is given by 
$
V_{KK}=  (\Gamma [0, \left({\ell}/{R}\right)^4 ]-
\frac{3}{2}) \hbar^2/2 N_B m_B R^2 \equiv U
$ \cite{ref25}, 
and the kelvon-fermion interaction coefficient is found to be $V_{KF}= {C_F R^2}/{2 N_B}$. Setting these coefficients equal to each other gives the third condition for supersymmetry, i.e.,
\begin{equation}
\frac{C_F}{C_B} =\left(\frac{\ell}{R} \right)^4 \left(\Gamma [0, \left({\ell}/{R}\right)^4 ]-
\frac{3}{2}\right),
\end{equation}
where $C_{B,F}$ are the spring constants associated with the radial confinement of the bosonic and fermionic atoms, respectively.

\begin{figure}
\includegraphics[scale=.63]{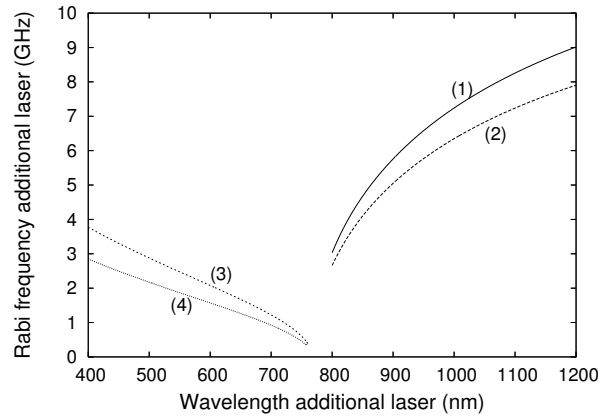}
\caption{Tuning of the additional laser to obtain supersymmetric interactions for 1000 bosonic atoms per lattice site. The Rabi frequency versus the wavelength of the running laser for different wavelengths of the lattice laser beam: (1) 400 nm, (2) 740 nm, (3) 850 nm, and (4) 1200 nm. See Fig. 2 for the corresponding Rabi frequencies of the lattice laser beam.
}
\end{figure}

For the $^{87}$Rb-$^{40}$K mixture, supersymmetric interactions require the ratio of the radial trapping potentials for the bosonic and fermionic atoms to be large, of the order of 100. This cannot be achieved by a magnetic potential, since the magnetic moment of the two species is almost the same. To overcome this problem, we propose to apply an extra running laser beam along the $z$-direction. The new laser beam does not influence the one-dimensional potential wells, but it does change the radial confinement and can therefore be used to tune the interaction terms equal. In principle this second laser also introduces interference terms, but they turn out to be negligible, since they oscillate faster than the atoms are able to follow. Hence, the intensities of the two lasers can simply be added. The detuning of the second laser has to be opposite to the detuning of the first laser. For the second laser, we can again independently choose both the wavelength and the Rabi frequency as shown in Fig. 3.  This can be used to minimize the atom loss due to the red-detuned laser, but it turns out that atom loss is always quite small anyway for reasonable system parameters: at worst the lifetime of the system is already 3 seconds.

Combining all these effects, our superstring is described by the supersymmetric hamiltonian
\begin{eqnarray} \label{hamiltonian}
\hat{\mathcal{H}} &=& - t \sum_{\langle i j \rangle} ( b_i^\dagger b_j + c_i^\dagger c_j)
\\ && + \sum_i (- \mu b_i^\dagger b_i -\mu c_i^\dagger c_i +
\frac{U}{2} b_i^\dagger b_i^\dagger b_i b_i+ U b_i^\dagger b_i
c_i^\dagger c_i). \nonumber
\end{eqnarray}
Here $b_i$ is the annihilation operator of a kelvon at site $i$, $c_i$ is the annihilation operator of a fermion at site $i$, and $\langle i j \rangle$  means that the summation runs over neighboring sites.
Note that supersymmetry means that 
$\lbrack \hat{\mathcal{Q}}, \hat{\mathcal{H}}\rbrack=0$, with $\hat{\mathcal{Q}}=\sum_i b_i c_i^\dagger$.
For $\Omega< \Omega_c$, the chemical potential is positive and the kelvons are thus
unstable towards Bose-Einstein
condensation, which physically corresponds to the tendency of the
superstring to spiral out of the center of the gas. The metastable superstring
at the center can be regarded as a system that is quenched deep into
an ordered phase, but is yet to move to equilibrium. The Bose-Einstein condensation of kelvons corresponds to a quantum phase transition that spontaneously
breaks supersymmetry, which manifests itself by the kelvon dispersion becoming
linear at long wavelengths.
Thus our proposed system naturally allows for a detailed study of the
nonequilibrium dynamics of supersymmetry breaking by monitoring the dynamics of the superstring.

\begin{figure}[t] 
\begin{minipage}[b]{.6cm} (a) \vspace{.4cm} \end{minipage} 
\includegraphics[scale=.3]{bubble.eps}
\hspace{.5cm}
\begin{minipage}[b]{.6cm} (b) \vspace{.4cm} \end{minipage} 
\includegraphics[scale=.3]{bubble_wigly.eps}
\caption{The second-order Feynman diagrams describing the dissipation
processes due to the creation of (a) additional pairs of bosonic and (b) fermionic excitations. The full lines correspond to bosons and the dotted lines to fermions. Because the right diagram contains a fermionic loop, it comes with an additional minus sign and the two diagrams thus cancel exactly in the supersymmetric case.}
\label{diagrams}
\end{figure}

It is important to realize that if supersymmetry is unbroken, i.e., the superstring is in a supersymmetric quantum state, the kelvon and fermion
modes should have the same average occupation number, i.e., $\langle b_i^\dagger b_i
\rangle = \langle c_i^\dagger c_i \rangle$.  Since the total number
of fermions is conserved, 
the supersymmetry in this case protects the superstring against dissipation, because the
spiraling out of the superstring would imply the creation of extra kelvons.
This results in a very long lifetime of the ultracold superstring, which can also be understood theoretically as
an exact cancellation of the
second-order Feynman diagrams describing the dissipation
processes due to the creation 
of additional pairs of bosonic and fermionic excitations, respectively, which are shown in Fig. \ref{diagrams}.
In addition, the equality of the average kelvon and fermion number
allows us to devise an experimental measure for the proximity to the
supersymmetric point.  The number of kelvons $\langle b_i^\dagger b_i \rangle$ can be obtained from the number of bosonic atoms $N_B$ and the 
mean-square displacement $\langle r^2 \rangle$ of the pancake vortices, which can be measured
by imaging along the
$z$-direction the size of the circle within which the
vortex positions are concentrated \cite{ref25}.  The average fermion number $\langle c_i^\dagger c_i\rangle=N_F$ can be determined by absorption measurements.  Hence also the quantity
$
( \langle b_i^\dagger b_i \rangle - \langle c_i^\dagger c_i \rangle )^2 =
(
 N_B { \langle r^2 \rangle}/{R^2} -  1/2- N_{F}
)^2
$
can be measured.  This quantity has an absolute minimum of zero at the
supersymmetric point, so that its magnitude is a measure of the
deviation from supersymmetry. The experimental precision that can be reached 
in approaching the supersymmetric point will mainly be limited by the uncertainty in the total number of fermions in the system. 
Nevertheless, the observation of the effects of supersymmetry in our setup should be in experimental reach with the existing technology. 


We are grateful for helpful discussions with Randy Hulet and Jan Ambj{\o}rn.

\end{document}